\documentclass[aps,prb,twocolumn,superscriptaddress,draft]{revtex4}
\input{epsf}
\usepackage{amsmath}
\usepackage{amssymb}
\input{epsf}

\newcommand{\rem}[1]{}

\begin{document}

\title{Quantum simulation of the single-particle Schr\"odinger equation}
\author{Giuliano Benenti}
\email{giuliano.benenti@uninsubria.it}
\affiliation{CNISM, CNR-INFM \& Center for Nonlinear and Complex Systems,
Universit\`a degli Studi dell'Insubria, Via Valleggio 11, 22100 Como, Italy}
\affiliation{Istituto Nazionale di Fisica Nucleare, Sezione di Milano,
via Celoria 16, 20133 Milano, Italy}
\author{Giuliano Strini}
\email{giuliano.strini@mi.infn.it}
\affiliation{Dipartimento di Fisica, Universit\`a degli Studi di Milano,
Via Celoria 16, 20133 Milano, Italy}
\date{\today}
\begin{abstract}
The working of a quantum computer is described in the concrete example 
of a quantum simulator of the single-particle Schr\"odinger equation.
We show that a register of $6-10$ qubits is sufficient to realize a 
useful quantum simulator capable of solving in an efficient way 
standard quantum mechanical problems.
\end{abstract}

\maketitle

\section{Introduction}

During the last years quantum information \cite{nielsen,qcbook} 
has become one of the new hot topic field in physics, 
with the potential to revolutionize many areas of science and technology.
Quantum information replaces the laws of classical physics,
applied to computation and communication, with the more fundamental 
laws of quantum mechanics. This becomes increasingly important 
due to technological progress reaching smaller and smaller scales.
Extrapolating the miniaturization process 
of integrated circuits one would estimate that we shall reach 
the atomic size for storing a single bit of information around
the year 2020 (at the present time the typical size of circuit
components is of the order of 100 nanometers). At that point, 
quantum effects will become unavoidably dominant. 

The final aim of quantum computation is to build a machine 
based on quantum logic, that is, a machine that can process the information
and perform logic operations in agreement with the laws of quantum mechanics.
In addition to its fundamental interest, a large scale quantum 
computer, if constructed, would advance computing power beyond 
the capabilities of classical computation. 

The purpose of the present paper is to illustrate the main features of 
a quantum computer in the concrete example of a quantum simulator 
of the single-particle Schr\"odinger equation. We wish to show 
that quantum logic, already with a small number of qubits, 
allows the efficient simulation of basic quantum mechanical problems.
A quantum simulator with $6-10$ qubits could produce a real 
picture book of quantum mechanics. We believe that snapshots 
from this book will help develop a physical intuition of the power 
of quantum computation. 

Our paper adds to other existing pedagogical presentations of various 
apects of quantum computation, including Shor's quantum algorithm for  
integer factorization \cite{gerjuoy}, Grover's quantum 
search algorithm \cite{grover,zhang}, quantum information processing
by means of nuclear magnetic resonance techniques \cite{havel},
using linear optics \cite{skaar},
with cold atoms and ions \cite{cirac}, or with superconducting 
circuits \cite{you}, 
quantum measurements \cite{brandt}, decoherence \cite{ford},
quantum error-correction and fault-tolerant
quantum computation \cite{preskill}.  

\section{Quantum logic}
\label{sec:qlogic}

The elementary unit of quantum information is the    
\emph{qubit} (the quantum
counterpart of the classical bit) and a quantum computer may be
viewed as a many-qubit system. Physically, a qubit is a
two-level system, like the two spin states of a spin-$\frac12$
particle, the polarization states of a single
photon or two states of an atom. 
 
A classical bit is a system that can exist in two distinct states, which are
used to represent $0$ and $1$, that is, a single binary digit. The only
possible operations (classical gates) in such a system are the identity
($0\to0$, $1\to1$) and NOT ($0\to1$, $1\to0$). 
In contrast, a quantum bit (qubit) is a two-level quantum system,
described by a two-dimensional complex Hilbert space. In this space, one
may choose a pair of normalized and mutually orthogonal quantum
states, called $|0\rangle$ and $|1\rangle$, 
to represent the values $0$ and $1$ of a 
classical bit. These two states form a computational basis.
From the \emph{superposition principle}, any state of the qubit 
may be written as
\begin{equation}
  |\psi\rangle = \alpha |0\rangle + \beta |1\rangle \,,
  \label{alphabetaqubit}
\end{equation}
where the amplitudes $\alpha$ and $\beta$ are complex numbers,
constrained by the normalization condition
$|\alpha|^2 + |\beta|^2 = 1$.
Moreover, state vectors are defined up to a global phase of no physical
significance. Therefore, $|\psi\rangle$ only depends on 
two real parameters and we may write
$|\psi\rangle=\cos\frac{\theta}{2}|0\rangle+
e^{i\phi}\sin\frac{\theta}{2}|1\rangle$, with
$0\le\theta\le\pi$ and $0\le\phi<2\pi$.

A collection of $n$
qubits is known as a \emph{quantum register} of size $n$. 
Its wave function resides in a $2^n$-dimensional
complex Hilbert space.
While the state of an $n$-bit classical computer is described in binary
notation by an integer $k\in \{0,1,...,2^n-1\}$,
\begin{equation}
  k = k_{n-1} \, 2^{n-1} + \cdots + k_1 \, 2 + k_0 \,,
  \label{binary}
\end{equation}
with $k_0,k_1,\dots ,k_{n-1}\in \{0,1\}$ binary digits, the state of an
$n$-qubit quantum computer is
\begin{equation}
  |\psi\rangle  =
  \sum_{k=0}^{2^n - 1} c_k \, |k\rangle,
  \label{superposition}
\end{equation}
where 
$|k\rangle \equiv |k_{n-1}\rangle \cdots
|k_1\rangle |k_0\rangle$, with $|k_j\rangle$ state of 
the $j$-th qubit, 
and $\sum_{k=0}^{2^n-1} |c_k|^2 = 1$.
Note that we use the shorthand notation 
$|k_{n-1}\rangle \cdots |k_1\rangle |k_0\rangle$ for the tensor product
$|k_{n-1}\rangle\otimes \cdots
\otimes |k_1\rangle \otimes |k_0\rangle$.
Taking into account the normalization condition and the 
fact that a quantum state is only defined up to an overall
phase, the state of an $n$-qubit quantum computer is determined
by $2\times 2^n -2$ independent real parameters.

The superposition principle is clearly visible in
Eq.~(\ref{superposition}): while $n$ classical bits can store only a single
integer $k$, the $n$-qubit quantum register can be prepared in the
corresponding state $|k\rangle$ of the computational basis, but also in a
superposition. We stress that the number of states of the computational basis
in this superposition can be as large as $2^n$, which grows
exponentially with the number of qubits. The superposition principle opens up
new possibilities for efficient computation. 
When we perform a computation on a classical
computer, different inputs require separate runs. In contrast, a
quantum computer can perform a computation for exponentially many inputs on a
single run. This huge parallelism is 
the basis of the power of quantum computation.

We stress that the superposition principle is not a uniquely quantum feature.
Indeed, classical waves satisfying the superposition
principle do exist. For instance, we may consider the wave
equation for a vibrating string with fixed endpoints.
It is therefore also important to point out the role of
\emph{entanglement} for the power of quantum computation,
as compared to any classical computation.
Entanglement is arguably the most spectacular and counter-intuitive
manifestation of quantum mechanics, observed in composite quantum systems: it
signifies the existence of non-local correlations between
measurements performed on particles who interacted in the past
but now are located arbitrarily far away. 
Mathematically, we say that a two-particle state $|\psi\rangle$
is entangled, or non-separable, if it cannot be written as a simple
product $|k_1\rangle|k_2\rangle$ of two states
describing the first and the second subsystem, respectively,
but only as a superposition of such states: 
$|\psi\rangle=\sum_{k_1,k_2}c_{k_1k_2}|k_1\rangle|k_2\rangle$.
For instance, the (Bell) state 
$\frac{1}{\sqrt{2}}(|00\rangle+|11\rangle)$ is entangled,
while the state  
$\frac{1}{\sqrt{2}}(|00\rangle+|10\rangle)$ is 
separable (\emph{i.e.}, not entangled), since we can write
it in the form $\frac{1}{\sqrt{2}}(|0\rangle+|1\rangle)|0\rangle$. 

There is no entanglement in classical physics. 
Therefore, in order to represent the superposition of $2^n$ levels
by means of classical waves, these levels must belong to the same system.
Indeed, classical states of separate systems can never be superposed.
The overall state space of a composite classical system is the 
Cartesian product of the individual state spaces of the subsystems, 
while in quantum mechanics it is the tensor product.
Thus, to represent the generic $n$-qubit state of 
(\ref{superposition}) by classical waves 
we need a single system with $2^n$ levels.
If $\Delta$ is the typical energy separation
between two consecutive levels, the amount of energy required for this
computation is given by $\Delta2^n$. Hence, the amount of physical
resources needed for the computation grows exponentially with $n$.
In contrast, due to entanglement, in quantum
physics a general superposition of $2^n$ levels may be represented by means of
$n$ qubits. Thus, the amount of physical resources (energy) grows only
linearly with $n$.

To implement a quantum computation, we must be able to control the
evolution in time of the many-qubit state describing the quantum
computer. As far as the coupling to the
environment may be neglected, this evolution is unitary and
governed by the Schr{\"o}dinger equation.
It is well known that a small set of elementary logic gates
allows the implementation of any complex computation
on a classical computer. This is very important: it means that, when we change
the problem, we do not need to modify our computer hardware. Fortunately,
the same property remains valid for a quantum computer.
It turns out that,
in the quantum circuit model, each unitary transformation acting on
a many-qubit system can be decomposed into (unitary) quantum gates acting
on a single qubit and a suitable (unitary) quantum gate acting on two qubits.
Any unitary operation on a single qubit can be
constructed using only Hadamard and phase-shift gates.
The Hadamard gate is defined as follows: it turns
$|0\rangle$ into $(|0\rangle + |1\rangle)/\sqrt{2}$ and
$|1\rangle$ into $(|0\rangle - |1\rangle)/\sqrt{2}$.
The phase-shift gate (of phase $\delta$) turns
$|0\rangle$ into $|0\rangle$ and $|1\rangle$ into
$e^{i\delta}|1\rangle$.
Since global phases have no physical meaning, the states of 
the computational basis, $|0\rangle$ and $|1\rangle$, are unchanged.
However, a generic single-qubit state 
$\alpha|0\rangle+\beta|1\rangle$ is mapped into
$\alpha|0\rangle+\beta e^{i\delta}|1\rangle$. Since relative 
phases are observable, the state of the qubit has been changed by the 
application of the phase-shift gate.
We can decompose a generic unitary transformation acting on
a many-qubit state into a sequence of Hadamard, phase-shift and
controlled-not (CNOT) gates, 
where CNOT is a two-qubit gate, defined as follows:
it turns $|00\rangle$ into $|00\rangle$,
$|01\rangle$ into $|01\rangle$,
$|10\rangle$ into $|11\rangle$ and
$|11\rangle$ into $|10\rangle$.
As in the classical XOR gate, the CNOT gate flips the state of the
second (target) qubit if the first (control) qubit is in the state
$|1\rangle$ and does nothing if the first qubit is in the state
$|0\rangle$.
Of course, the CNOT gate, in contrast to the
classical XOR gate, can also be applied to any superposition of the
computational basis states.

\begin{figure}
\centerline{\epsfxsize=8.5cm\epsffile{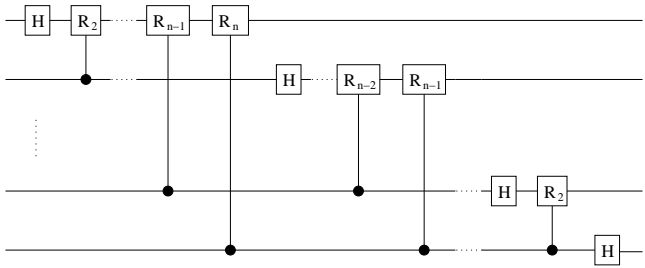}}
\caption{A quantum circuit implementing the quantum Fourier transform
(except for the fact that the order of qubits in the output 
is reversed, so that
one has simply to relabel qubits in the output state). As usual in
the graphical representation of quantum circuits, each line corresponds 
to a qubit and any sequence of logic gates must be read from the left
(input) to the right (output). 
From bottom to top, qubits run from 
the least significant ($k_0$, according to binary notation (\ref{binary}))
to the most significant ($k_{n-1}$). 
Here a qubit is said to be more significant than another if its flip gives 
a larger variation in the integer number coded by the state of the
$n$ qubits.
A square with $H$ written inside stands for the Hadamard gate, 
while for controlled-phase shift gates 
CPHASE$\left(\frac{2\pi}{2^k}\right)$ gates 
we draw a circle on the control qubit and a square with $R_k$ 
written inside on the target qubit.
Note that to implement the inverse Fourier transform it is sufficient
to run the circuit in this figure from right to left and with $R_k^\dagger$
instead of $R_k$. This is due to the facts that the Hadamard gate is 
self-inverse and $R_k^\dagger=R_k^{-1}$.}
\label{fig:qft}
\end{figure}

The decomposition of a generic unitary transformation of an $n$-qubit
system into elementary quantum gates is in general inefficient,
that is, it requires a number of gates exponentially large
in $n$ (more precisely, $O(n^2 4^n)$ quantum gates).
However, there are unitary transformations that can
be computed efficiently in the quantum circuit model, namely
by means of a number of elementary gates polynomial in $n$.
This is the case in many computational problems of interest. 
A very important example is given by the quantum Fourier transform,
mapping a generic $n$-qubit state $\sum_{k=0}^{N-1} f(k) |k\rangle$
into $\sum_{l=0}^{N-1} \tilde{f}(l) |l\rangle$, where 
$N=2^n$ and the vector
$\{\tilde{f}(0),...,\tilde{f}({N-1})\}$ is 
the discrete Fourier transform of the vector
$\{f(0),...,f({N-1})\}$, that is,
$\tilde{f}(l)=\frac{1}{\sqrt{N}}
\sum_{k=0}^{N-1} e^{2\pi i k l/ N} f(k)$. It can be shown 
that this transformation can be efficiently implemented 
in $O(n^2=(\log_2 N)^2)$ 
elementary quantum gates, whereas the best known classical
algorithm to simulate the Fourier transform, the fast Fourier transform,
requires $O(N \log_2 N)$ elementary operations.
A quantum circuit computing the quantum Fourier transform is shown 
in Fig.~\ref{fig:qft}: it requires $n$ Hadamard and $n(n-1)/2$ 
controlled phase-shift gates. By definition, the controlled 
phase-shift gate CPHASE$(\delta)$ applies a phase shift 
$\delta$ to the target qubit only when the control qubit is in the 
state $|1\rangle$:
it turns $|00\rangle$ into $|00\rangle$,
$|01\rangle$ into $|01\rangle$,
$|10\rangle$ into $|10\rangle$ and
$|11\rangle$ into $e^{i\delta}|11\rangle$.
The quantum Fourier transform is an essential subroutine in many quantum
algorithms, including the quantum simulator described in this paper.

\section{Quantum algorithms}

As we have seen, the power of quantum computation is due to the inherent
\emph{quantum parallelism} associated with the superposition principle.
In simple terms, a quantum computer can process a large number of
classical inputs in a single run.
For instance, starting from the input state
$\sum_{k=0}^{2^n-1}c_k|k\rangle |0\rangle$,
we may obtain the output
state
\begin{equation}
\sum_{k=0}^{2^n-1}c_k|k\rangle |f(k)\rangle.
\label{fsuperposition}
\end{equation}
Therefore, we have computed the function $f(k)$ for all $k$ in a
single run. Note that a quantum computer implements reversible 
unitary transformations and that we need two quantum registers 
in (\ref{fsuperposition}) 
to compute in a reversible way a generic function $f$.
We emphasize that it is not an easy task to extract useful information
from the output state. The problem is that this information
is, in a sense, hidden.
Any quantum computation ends up with a projective measurement
in the computational basis: we measure the
qubit polarization ($0$ or $1$) for all the qubits.
The output of the
measurement process is inherently probabilistic and the probabilities of
the different possible outputs are set by the basic postulates of quantum
mechanics. Given the state (\ref{fsuperposition}), we obtain
outcome $|\bar{k}\rangle|f(\bar{k})\rangle$ with probability $|c_{\bar{k}}|^2$.
Hence, we end up with the evaluation of the function $f(k)$ for a single
$k=\bar{k}$, exactly as with a classical computer. Nevertheless, there exist
quantum algorithms that exploit \emph{quantum
interference} to efficiently extract useful information.
Note that in principle one could consider more general, non projective
measurements. However, this would not modify considerations on
the efficiency of a given quantum algorithm. 
The reason is that generalized mesurements, after the addition of 
ancillary qubits, can always be represented by unitary transformations 
followed by standard projective measurements \cite{nielsen,qcbook}.

In 1994, Shor discovered a quantum algorithm which
factorizes large integers exponentially faster than any
known classical algorithm. It was also shown by Grover 
that the search of an item in an unstructured database can be done 
with a square root speed up over any classical algorithm.
A third relevant class of quantum algorithms is the simulation
of physical systems. In particular, the power of a quantum computer 
is intuitive for quantum many-body problems. It is indeed well known that 
the simulation of such problems on a classical computer is 
a difficult task as the size of the Hilbert space grows exponentially 
with the number of particles. For instance, if we wish to simulate a chain 
of $n$ spin-$\frac12$ particles, the size of the Hilbert space is $2^n$. 
Namely, the state of this system is determined by $2^n$ complex numbers. 
As observed by Feynman in the 1980's, the growth in memory requirement
is only linear on a quantum computer, which is itself a many-body quantum 
system. For example, to simulate $n$ spin-$\frac12$ particles we 
only need $n$ qubits. Therefore, a quantum computer operating with only 
a few tens of qubits can outperform a classical computer. Of course, this 
is only true if we can find efficient quantum algorithms 
to extract useful information from the quantum computer.
Quite interestingly, it has been shown that a quantum computer 
can be useful not only for the investigation of the properties 
of many-body quantum systems, but also for the study of the quantum 
and classical dynamics of complex single-particle systems.
As we shall see in the next section, basic quantum mechanical 
problems already can be simulated with $6-10$ qubits, while about
$40$ qubits would allow one to make computations
inaccessible to today's supercomputers. We note that this figure has
to be compared with the more than $1000$ qubits required to the Shor's 
algorithm to outperform classical computations. Moreover, these estimates 
neglect the computational overhead needed to correct errors
and error-correction becomes necessary to obtain reliable results
by means of a large-scale quantum computer.  

\section{The quantum simulator}
\label{sec:simulator}

To illustrate the working of a quantum algorithm by means of a
concrete example, let us consider the quantum-mechanical motion of a
particle in one dimension (the extension to 
higher dimensions is straightforward) \cite{wiesner,zalka,strini}.
It is governed by the Schr{\"o}dinger equation
\begin{equation}
  i \hbar \, \frac{d}{dt} \, \psi(x,t) =
  H \, \psi(x,t) \,,
  \label{1Dsch}
\end{equation}
where the Hamiltonian $H$ is given by
\begin{equation}
  H =
  H_0 + V(x) =
  -\frac{\hbar^2}{2m} \, \frac{d^2}{dx^2} + V(x) \,.
\end{equation}
The Hamiltonian $H_0=-(\hbar^2\!/2m)\,d^2\!/dx^2$ governs the free motion of
the particle, while $V(x)$ is a one-dimensional potential. To solve
Eq.~(\ref{1Dsch}) on a quantum computer with finite resources (a finite number
of qubits and a finite sequence of quantum gates), we must first of all
discretize the continuous variables $x$ and $t$.
If the motion essentially
takes place inside a finite region, say $-d\le{x}\le{d}$, we decompose this
region into $2^n$ intervals of length $\Delta=2d/2^n$ and
represent these intervals by means of the Hilbert space of an $n$-qubit
quantum register (this means that the discretization step drops exponentially
with the number of qubits). Hence, the wave function $\psi(x,t)$ is
approximated by 
\begin{equation}
  \sum_{k=0}^{2^n-1} c_k(t) \, |k\rangle =
  \frac1{\mathcal{N}} \sum_{k=0}^{2^{n}-1} \psi(x_k,t) \, |k\rangle \,,
\label{psidiscretized}
\end{equation}
where
\begin{equation}
  x_k \equiv -d + \left(k + \tfrac12\right) \Delta,
\label{xdiscretized}
\end{equation}
$|k\rangle=|k_{n-1}\rangle |k_{n-2}\rangle\dots |k_0\rangle$
is a state of the computational basis of the $n$-qubit quantum register
and
\begin{equation}
  {\mathcal{N}} \equiv \sqrt{\sum_{k=0}^{2^n-1} |\psi(x_k,t)|^2}
\label{discnorm}
\end{equation}
is a factor that ensures correct normalization of the wave
function. It is intuitive that (\ref{psidiscretized}) provides a good
approximation to $|\psi\rangle$ when the discretization step $\Delta$
is much smaller than the shortest length scale relevant for the 
description of the system.

The Schr{\"o}dinger equation
(\ref{1Dsch}) may be integrated formally by propagating the
initial wave function $\psi(x,0)$ for each time-step $\epsilon$ as
follows:
\begin{equation}
  \psi (x,t + \epsilon) =
  e^{-\frac{i}\hbar [H_0 + V(x)] \epsilon} \, \psi(x,t) \,.
\end{equation}
If the time-step $\epsilon$ is small enough
(that is, much smaller than the time scales of interest for the 
dynamics of the system), it is possible to write
\begin{equation}
  e^{-\frac{i}\hbar [H_0+ V(x)] \, \epsilon} \approx
  e^{-\frac{i}\hbar H_0 \, \epsilon}
  e^{-\frac{i}\hbar V(x) \, \epsilon} \,.
  \label{Trotter}
\end{equation}
Note that this equation, known as the Trotter decomposition, is
only exact up to terms of order $\epsilon^2$ since the operators
$H_0$ and $V$ do not commute. The operator on the right-hand side of
Eq.~(\ref{Trotter}) is still unitary and simpler than that on the
left-hand side.
We can now take advantage of the fact that
the Fourier transform can be
efficiently performed by a quantum computer. We call $p$ the momentum variable
conjugate to $x$, that is, $-i\hbar (d/dx)=F^{-1}p F$, where $F$ is the
Fourier transform. Therefore, we can write the first operator in the
right-hand side of (\ref{Trotter}) as
\begin{equation}
  e^{-\frac{i}\hbar H_0 \, \epsilon} =
  F^{-1} \,
  e^{-\frac{i}\hbar \left(\frac{p^2}{2m}\right) \epsilon} \,
  F \,.
\end{equation}
In this expression, we pass, by means of the Fourier transform $F$,
from the $x$-representation to the $p$-representation, in which this operator
is diagonal. Then, using the inverse Fourier transform $F^{-1}$, we return to
the $x$-representation, in which the operator $\exp(-iV(x)\epsilon/\hbar)$ is
diagonal. The wave function $\psi(x,t)$ at time $t=l\epsilon$ is 
obtained (up to errors $O(\epsilon^2)$) from the initial wave function 
$\psi(x,0)$ by applying $l$ times the unitary operator
\begin{equation}
  F^{-1} \,
  e^{-\frac{i}\hbar \left(\frac{p^2}{2m}\right) \epsilon} \,
  F \,
  e^{-\frac{i}\hbar V(x) \, \epsilon}.
\end{equation}

\begin{figure}
\centerline{\epsfxsize=5cm\epsffile{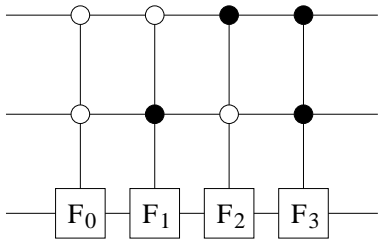}}
\caption{A quantum circuit implementing a generic diagonal 
unitary transformation (\ref{eifx}), for the case of $n=3$
qubits. Empty or full circles indicate that the operation on the
target qubit is active only when the control qubits are set to $0$ 
or $1$, respectively.}
\label{fig:diagonalU}
\end{figure}

Therefore, simulation of the Schr{\"o}dinger equation is now
reduced to the implementation of the Fourier transform plus diagonal operators
of the form
\begin{equation}
  |x\rangle \to e^{if(x)} \, |x\rangle.
  \label{eifx}
\end{equation}
Note that an operator of the form (\ref{eifx})
appears both in the computation of $\exp(-iV(x)\epsilon/\hbar)$ and of
$\exp(-iH_0\epsilon/\hbar)$, when this latter operator is written in the
$p$-representation. 
Any operator of the form (\ref{eifx}) can be implemented by means of 
$2^{n}/2$ generalized controlled-phase shift gates, which apply
the transformation $F_k$ to the target qubit only when the 
$n-1$ (control) qubits are in the state $|k\rangle$. Here $F_k$ is 
a single-qubit gate, mapping $|0\rangle$ into $e^{if(2k)}|0\rangle$ and 
$|1\rangle$ into $e^{if(2k+1)}|1\rangle$.
It is easy to check this construction for the three-qubit case shown in 
Fig.~\ref{fig:diagonalU}.
$F_0$ acts only when the
first two qubits are in the state $|00\rangle$ and therefore it sets the
phases $e^{if(0)}$ and $e^{if(1)}$ in front of the basis
vectors $|000\rangle$ and $|001\rangle$, respectively. Similarly,
$F_1$ acts only when the first two qubits are in the state
$|01\rangle$ and therefore it sets the phase $e^{if(2)}$ and
$e^{if(3)}$ in front of the basis vectors $|010\rangle$ and
$|011\rangle$ and so on.

Note that the implementation described in Fig.~\ref{fig:diagonalU}
is inefficient, as it scales exponentially with the number of qubits.
On the other hand, efficient (that its, polynomial in $n$) implementations
are possible for most of the cases of physical interest.
For instance, $n^2$ two-qubit phase-shift gates are sufficient
for the harmonic oscillator potential 
$V(x)=\frac{1}{2}m\omega^2x^2$.
Using Eqs. (\ref{xdiscretized}) and (\ref{binary}), we can write the 
discretized variable $x$ as $\alpha\sum_{j=0}^{n-1} (k_j 2^j +\beta)$, 
with the constants $\alpha=\Delta$ and 
$\beta=\frac{-d+\Delta/2}{\alpha n}$.
Therefore, $x^2=\alpha^2\sum_{j,l=0}^{n-1}
(k_j 2^j +\beta)(k_l 2^l +\beta)$ and finally
\begin{equation}
e^{-\frac{i}{\hbar}V(x)\epsilon}=
\prod_{j,l=0}^{n-1}e^{-i\gamma 
(k_j 2^j +\beta)(k_l 2^l +\beta)},
\end{equation}
where $\gamma=
m\omega^2\alpha^2\epsilon/(2\hbar)$. This is the product of 
$n^2$ phase-shift gates, each acting non-trivially (differently from 
identity) only on the qubits $j$ and $l$.
Since the kinetic energy $H_0$ is proportional to $p^2$, an 
analogous decomposition is readily obtained, in the momentum 
eigenbasis, for $\exp(-iH_0\epsilon/\hbar)$.
Efficient implementations are possible for piecewise analytic
potentials $V(x)$ but require, in general, the use of ancillary qubits.

Finally, we point out that 
there is an exponential advantage in memory requirements
with respect to classical computation,
While a classical computer needs $O(N)$ bits to load the state vector 
of a system of size $N$ (that is, the coefficients $\psi(x_k,t)$ 
of its expansion over the computational basis), a quantum computer
accomplishes the same task with just $n=\log_2 N$ qubits, namely
with memory resources only logarithmic in the system size. 
As we have discussed in Sec.~\ref{sec:qlogic}, this is possible 
thanks to entanglement of the qubits in quantum computer. 

\section{A few snapshots}

\begin{figure}
\centerline{\epsfxsize=8.5cm\epsffile{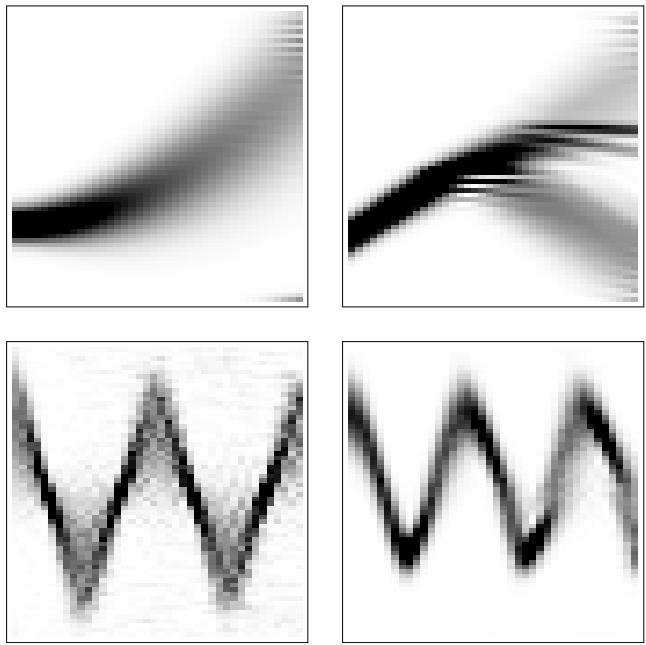}}
\caption{Plots of $|\psi(x,t)|^2$ 
with $n=6$ qubits 
($t$ horizontal axis, divided in $40$ time steps; 
$x$ vertical axis, discretized by means of a grid of $2^n=64$ points):
accelerated particle (top left),
transmission and reflection through a square barrier (top right),
harmonic potential, with a squeezed state of initial width 
twice that of a coherent state (bottom left),
anharmonic potential (bottom right). The initial state
$\psi(x,0)$ in all cases is a Gaussian wave packet.}
\label{fig:snapshots1}
\end{figure}

To illustrate the working of the quantum simulator, we 
simulate it on a classical computer,
obviously with an exponential slowing down with respect to
a true quantum computation.
Let us consider a few significant examples of single-particle quantum 
mechanical problems, for which plots of $|\psi(x,t)|^2$ are shown in 
Fig.~\ref{fig:snapshots1}. 
In contrast to classical simulations, 
in quantum computation we cannot access the wave function 
$\psi(x,t)$ after a single run up to time $t$. 
Indeed, each run is followed by a standard projective measurement 
on the computational basis, giving outcome $x_k$
with probability $|\langle k | \psi(x,t)\rangle|^2$.
Therefore, the probability distribution $|\psi(x,t)|^2$ 
may be reconstructed, up to statistical errors, only 
if the quantum simulation is repeated many times. 
If outcome $x_k$ is obtained $N_k$ times in $N$ runs, we can estimate
$|\psi(x_k,t)|^2$ as $\mathcal{N}^2 \frac{N_k}{N}$, with $\mathcal{N}$
normalization factor defined in Eq.~(\ref{discnorm}).

\begin{figure}
\centerline{\epsfxsize=6.cm\epsffile{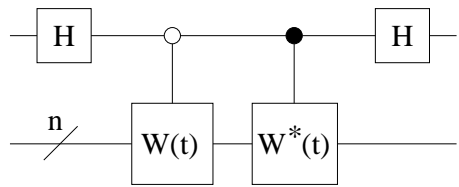}}
\caption{Quantum circuit measuring the squares of the 
real and imaginary parts of the wave function $\psi(x,t)$.
The line with a dash represents a set of $n$ qubits.}
\label{fig:ramsey}
\end{figure}

Fig.~\ref{fig:snapshots1} exhibits several interesting features 
of quantum mechanics. In the top left plot we can see the spreading
(quadratic in time) of a Gaussian wave packet for an accelerated 
particle. 
In the top right plot interference fringes appear when a 
Gaussian packet impinges on a square barrier.  
The bottom left plot shows the width oscillations for a squeezed 
state, back and forth between a minimum and a maximum value.
Finally, the bottom right plot illustrates the motion of a wave 
packet in a potential harmonic for $x>0$ and anharmonic 
($V(x)=\alpha x^3$) for $x<0$. We can see the deformation and
spreading of the wave packet when it moves in the anharmonic 
part of the potential. As a result, the initial coherence of the 
state is lost in a couple of oscillation periods. 

As shown in Fig.~\ref{fig:ramsey}, 
it is also possible to measure the squares of the real 
and imaginary parts of $\psi(x,t)$ by
means of a Ramsey type quantum interferometry method. For this purpose, 
we need a single ancillary qubit, initially in the state $|0\rangle$. 
The input state of the other $n$ qubits is $|0\rangle=|00\cdots0\rangle$ 
and they end up,
depending on the $|0\rangle$ or $|1\rangle$ state of the ancillary qubit, 
in the states $|\psi(t)\rangle=W(t)|0\rangle$ 
or $|\psi^\star(t)\rangle=W^\star(t)|0\rangle$. 
Note that the operator $W(t)=U(t)S$ includes both the state preparation
($|\psi(0)\rangle=S|0\rangle$) and the time evolution 
($|\psi(t)\rangle=U(t)|\psi(0)\rangle$).
Finally, a Hadamard gate applied to the ancillary qubit leads to
the $(n+1)$-qubit output state
\begin{equation}
|\Phi(t)\rangle=\frac{1}{2}\,|0\rangle(|\psi(t)\rangle+|\psi^\star(t)\rangle)+
\frac{1}{2}\,|1\rangle(|\psi(t)\rangle-|\psi^\star(t)\rangle).
\end{equation}
Hence, the probabilities of obtaining outcomes $0$ or $1$
from the measurement of 
the ancillary qubit and $x_k$ from the measurement 
of the other $n$ qubits are given by
\begin{eqnarray} 
P_0(x_k,t)&=|\langle 0 |\langle k |\Phi(t)\rangle|^2
=\left\{{\mathop\text{Re}}\left[\psi(x_k,t)\right]\right\}^2,
\\
P_1(x_k,t)&=|\langle 1 |\langle k |\Phi(t)\rangle|^2
=\left\{{\mathop\text{Im}}\left[\psi(x_k,t)\right]\right\}^2.
\end{eqnarray}

\begin{figure}
\centerline{\epsfxsize=8.5cm\epsffile{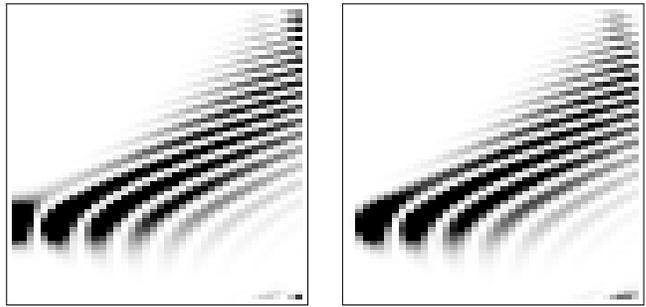}}
\caption{Plots of $\{{\mathop\text{Re}}[\psi(x,t)]\}^2$ and 
$\{{\mathop\text{Im}}[\psi(x,t)]\}^2$ for the uniformly accelerated particle
of Fig.~\ref{fig:snapshots1} top left.}
\label{fig:snapshots2}
\end{figure}

From the measurement of $\{{\mathop\text{Re}}\left[\psi(x_k,t)\right]\}^2$
and $\{{\mathop\text{Im}}\left[\psi(x_k,t)\right]\}^2$
we can derive the phase velocity from the isophase curves of the wave function
$\psi(x,t)=|\psi(x,t)|e^{i\theta(x,t)}$, corresponding to 
$\theta(x,t)=0,\pi$ (implying ${\mathop\text{Im}}\left[\psi(x_k,t)\right]=0$)
and $\theta(x,t)=\frac{1}{2}\pi,\frac{3}{2}\pi$ 
(implying ${\mathop\text{Re}}\left[\psi(x_k,t)\right]=0$).
The phase velocity is just given by the slope of the isophase curves.
In Fig.~\ref{fig:snapshots2}, 
$\{{\mathop\text{Re}}\left[\psi(x_k,t)\right]\}^2$ and 
$\{{\mathop\text{Im}}\left[\psi(x_k,t)\right]\}^2$
are shown for the quantum motion of a uniformly accelerated particle.
It is interesting to note that in this example the phase
velocity has strong local variations: it can change sign and also diverge.

\begin{figure}
\centerline{\epsfxsize=8.5cm\epsffile{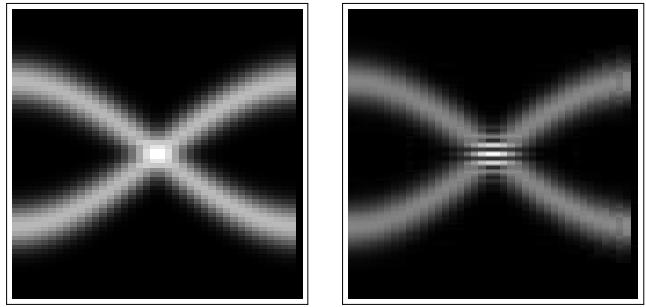}}
\caption{Plots of $|\psi(x,t)|$ for the superposition of two
counterpropagating wave packets in a harmonic potential, 
simulated with $n=6$ (left) and $n=8$ (right) qubits.}
\label{fig:snapshots3}
\end{figure}

The huge memory capabilities of the quantum computer appear in
the plots of Fig.~\ref{fig:snapshots3}, where the dynamics 
of the superposition of two counterpropagating Gaussian packets
in a harmonic potential is considered. When the two packets 
collide interference fringes appear, similarly to the double-slit
experiment for free particles. These fringes are hardly visible 
with $n=6$ qubits but already very clear with $n=8$ qubits.
Since the number of levels grows exponentially with the number of
qubits, the discretization step reduces exponentially.
Therefore, position resolution improves $4$ times when 
moving from $6$ to $8$ qubits (of course, there is a lowest
resolution limit imposed by the Heisenberg uncertainty principle).

Finally, we point out that the use of the quantum Fourier transform
in the simulation of the Schr\"odinger equation automatically
imposes periodic boundary conditions. Such boundary effects become
relevant when the finite region $-d\le x \le d$ is not large enough 
for a correct description of the motion. This problem can be seen in the 
two top plots of Fig.~\ref{fig:snapshots1}. In the case of the uniformly 
accelerated particle, when the wave packet gets close to the upper
border $x=+d$, there exists a spurious non-negligible probability to find the 
particle close to the lower border $x=-d$. With regard to the problem of 
the scattering of a particle by a square potential, we observe interference
fringes between the transmitted and the reflected packets, when both 
are close to the boundaries.   
In general, it might be interesting for several physical problems to
consider non-periodic boundary conditions. Keeping the quantum Fourier
structure of the problem, this can be done if suitable fictitious potentials
are added. For instance, zero boundary conditions may be approximated 
if a very high, in practice impenetrable potential barrier encloses 
the region of interest for the motion. Twisted boundary conditions
[$\psi(d,t)=e^{i\phi}\psi(-d,t)$] are obtained by adding a suitable
Aharonov-Bohm flux $\phi$ to the problem. 

\section{Final remarks}

To asses the feasibility of a quantum computer, one should
take into account that, in any realistic implementation,
errors due to imperfections in the quantum computer hardware or  
to the undesired computer-environment coupling unavoidably appear.
First studies \cite{strini,qcbook} have shown that a quantum simulator
with $n=6-10$ qubits is quite robust against errors.
Nowadays, these numbers of qubits are available in experiments with 
liquid state nuclear magnetic resonance-based quantum processors 
and with cold ions in a trap \cite{qcbook}. On the other hand, the number
of elementary quantum gates required for the simulations 
discussed in Figs.~\ref{fig:snapshots1}, \ref{fig:snapshots2}, and
\ref{fig:snapshots3} is $O(10^4)$, much larger than 
the sequences of $10-100$ gates
that can be reliably implemented in present-day laboratory experiments. 

Even though the time when a useful quantum computer will be realized is 
uncertain, quantum computation is a very promising and fascinating 
field of investigation in physics, mathematics and computer science.
Furthermore, we believe that quantum computation and more generally 
quantum information provide an excellent approach 
to teaching basic quantum mechanics, because they deal in an intuitive,
appealing and mathematically simple way with the main features 
of the quantum theory, from the superposition principle to 
entanglement, quantum interference and quantum measurements. 

One of us (G.B.) acknowledges support by MIUR-PRIN 2005 (2005025204).


\begin{thebibliography}{9}

\bibitem{qcbook} G. Benenti, G. Casati and G. Strini,
\textit{Principles of Quantum Computation and Information},
Vol. I: Basic concepts (World Scientific, Singapore, 2004);
Vol. II: Basic tools and special topics
(World Scientific, Singapore, 2007).

\bibitem{nielsen} M.A. Nielsen and I.L. Chuang,
\textit{Quantum computation and quantum information}
(Cambridge University Press, Cambridge, 2000).

\bibitem{gerjuoy} E. Gerjuoy, ``Shor's factoring algorithm
and modern cryptography. An illustration of the capabilities 
inherent in quantum computers'', Am J. Phys. \textbf{73}, 521--540 (2005).

\bibitem{grover} L.K. Grover, ``From Schr\"odinger equation to the quantum 
search algorithm'', Am. J. Phys. \textbf{69}, 769--777 (2001).

\bibitem{zhang} J. Zhang and Z. Lu,
``Similarity between Grover's quantum search 
algorithm and classical two-body collisions''
Am. J. Phys. \textbf{71}, 83--86 (2003).

\bibitem{havel} T.F. Havel, D.G. Cory, S. Lloyd, N. Boulant, E.M. Fortunato,
M.A. Pravia, G. Teklemarian, Y.S. Weinstein, A. Bhattacharyya, and J. Hou,
``Quantum information processing by nuclear magnetic resonance spectroscopy'',
Am. J. Phys. \textbf{70}, 345--362 (2002).

\bibitem{skaar} J. Skaar, J.C.G. Escart\'/i/n, and H. Landro, 
``Quantum mechanical description of linear optics'',
Am. J. Phys. \textbf{72}, 1385--1391 (1004).

\bibitem{cirac} J.I. Cirac and P. Zoller, ``New frontiers in quantum 
information with atoms and ions'', Phys. Today \textbf{57} (3), 38--44 (2004). 

\bibitem{you} J.Q. You and F. Nori, ``Superconducting circuits and quantum 
information'', Phys. Today \textbf{58} (11), 42--47 (2005).

\bibitem{brandt} H.E. Brandt,
``Positive operator valued measure in quantum information processing'',
Am. J. Phys. \textbf{67}, 434--439 (1999).

\bibitem{ford} G.W. Ford and R.F. O'Connell,
``Wave packet spreading: Temperature and squeezing effects with 
applications to quantum measurement and decoherence'',
Am J. Phys. \textbf{70}, 319--324 (2002).    

\bibitem{preskill} J. Preskill, ``Battling decoherence: 
The fault-tolerant quantum computer'', Phys. Today \textbf{52} (6), 
24-30 (1999).

\bibitem{wiesner}
S. Wiesner, ``Simulation of many-body quantum systems by a quantum computer'',
quant-ph/9603028.

\bibitem{zalka}
C. Zalka, ``Efficient simulation of quantum systems by quantum computers'',
Fortschr. Phys. \textbf{46}, 877--879 (1998).

\bibitem{strini}
G. Strini, ``Error sensitivity of a quantum simulator I: a first example'',
Fortschr. Phys. \textbf{50}, 171--183 (2002).

\end{thebibliography}
\end{document}